\documentclass[a4paper,11pt]{article}
\pdfoutput=1 

\usepackage{jheppub} 

\usepackage{fancyhdr}
\usepackage{graphicx}

\usepackage{braket}

\newcommand{\be}{\begin{equation}}
\newcommand{\ee}{\end{equation}}
\newcommand{\bea}{\begin{eqnarray}}

\newcommand{\eea}{\end{eqnarray}}

\newcommand{\refe}[1]{Eqn.~(\ref{#1})}

\newcommand{\Rmnum}[1]{\expandafter\@slowromancap\romannumeral #1@}

 \title{\boldmath Natural Supersymmetry and Dynamical Flavour with Meta-stable Vacua}

\author[\diamondsuit]{Steven Abel}
\author[\dagger]{and Moritz McGarrie}

\affiliation[\diamondsuit]{Institute for Particle Physics Phenomenology,
Durham University, South Road, Durham, DH1 3LE, UK}
\affiliation[\dagger]{School of Physics and Centre for Theoretical Physics,
University of the Witwatersrand, Johannesburg, WITS 2050, South Africa}

\emailAdd{s.a.abel@durham.ac.uk}
\emailAdd{moritz.mcgarrie@wits.ac.za}

\abstract{We show how gauged flavour breaking and $\mathcal{N}=1$ supersymmetry breaking can be dynamically aligned to produce natural models. Supersymmetry is broken in a metastable vacuum, while a weakly gauged flavour symmetry is identified with an $SU(3)_F$ subgroup of the global symmetry of the SQCD model. We find that alignment can easily occur either through strongly coupled flavour models, with 
SQCD bound states playing the role of flavons, or entirely weakly coupled and renormalizable models with elementary $SU(3)_F$ adjoint flavons. 
 In both cases it is essential that  
the $SU(3)_F$ breaking, and hence all the flavour structure, is driven dynamically by the SUSY breaking. The resulting  flavour gauge mediation 
in conjunction with the usual gauge mediation contribution leads to naturally light third generations squarks, with first and second generations above exclusions.
}

\begin{document} 
\maketitle
\flushbottom

\section{Introduction} \label{sec:intro}

In its original format, gauge mediation had the great advantage that flavour universality provided an explanation for the absence of flavour changing neutral currents. However flavour universality (and gauge mediation in general) is becoming less attractive, because at the time of writing the LHC has excluded first and second generation squarks in the TeV range \cite{ATLASsqgl,CMSsqgl}, while naturalness of the Higgs sector (and the much weaker bounds on 3rd generation squarks of around 300-400 GeV \cite{ATLAS3rd,CMS3rd}), favours a sub-TeV 3rd generation.

 Such large splittings must be imprinted in the high scale boundary conditions of the soft mass-terms in an inverted hierarchy or effective SUSY model. This observation was the starting point of a number of models of flavour; see for instance \cite{Shadmi:2011hs,Kang:2012ra,Albaid:2012qk,Abdullah:2012tq,Craig:2012di,Auzzi:2012dv,Bharucha:2013ela,Dudas:2013pja,Calibbi:2013mka,Galon:2013jba}.

An inverted hierarchy suggests that the dynamics of supersymmetry breaking are, at least in part, related to the dynamics of flavour. Recently ref.\cite{Brummer:2013upa} proposed that the dynamics of (weakly gauged) flavour breaking $SU(3)_F\rightarrow SU(2)_F$ are actually \emph{intimately} related to supersymmetry breaking, with some of the supersymmetry breaking fields being charged under $SU(3)_F$. This results in gauge (vector multiplet) messengers \cite{Intriligator:2010be}, and can generate the required inverted hierarchies amongst the soft masses. At the messenger scale, integrating out the gauge (vector multiplet) messengers generates negative mass-squared terms for stops, which after renormalisation group running obtain positive but small values at the electroweak scale.  Such boundary conditions are strikingly natural as they significantly reduce the radiative effects of the stop soft term on $\delta m_{H_u}^2$.  

In this work we wish to show that such alignment can be readily generated dynamically within an $N_f$-flavour ISS type configuration, with an $SU(3)_F$ subgroup of the global flavour symmetry, namely $$ SU(N_f)_L\times SU(N_f)_R\supset  SU(3)_L\times SU(3)_R \supset [SU(3)_L\times SU(3)_R]_{\rm V}  = SU(3)_F$$ becoming the weakly gauged SM flavour symmetry. The remaining (anomalous) global symmetry, namely $ SU(N_f-3)_L\times SU(N_f-3)_R \times [SU(3)_L\times SU(3)_R]_{\rm A}$ can be explicitly broken to avoid physical Goldstone modes, while the spontaneous breaking of $SU(3)_F$ induced by the ISS mechanism gives calculable models with inverted hierarchies. 

Although the discussion concerns flavour, we would like to avoid making very restrictive assumptions about how the hierarchical structure comes about. Therefore we will consider both strong and weak models of flavour. In both cases the flavour symmetry is weakly gauged, but in the former case one generally supposes that the flavour sector comprises bound states of some SQCD theory \cite{Seiberg:1994pq,Intriligator:1995au}. The flavour hierarchies would then be governed by the dynamical scale(s) of that theory. In the context of ISS one obviously works in the  perturbative (magnetic dual) description. The weak model of flavour by contrast requires elementary flavon fields that merely couple to the ISS sector. Here we shall assume these states to transform in the adjoint of flavour. While adjoints are not normally considered as flavons, it is natural in the context of ISS (because of the bound-state mesons), and bring advantages of their own to do with the strong CP problem, as discussed in \cite{Masiero:1998yi,Abel:2000hn,Abel:2001ur}.


\begin{figure}
 \centering
 \includegraphics[width=.6\textwidth]{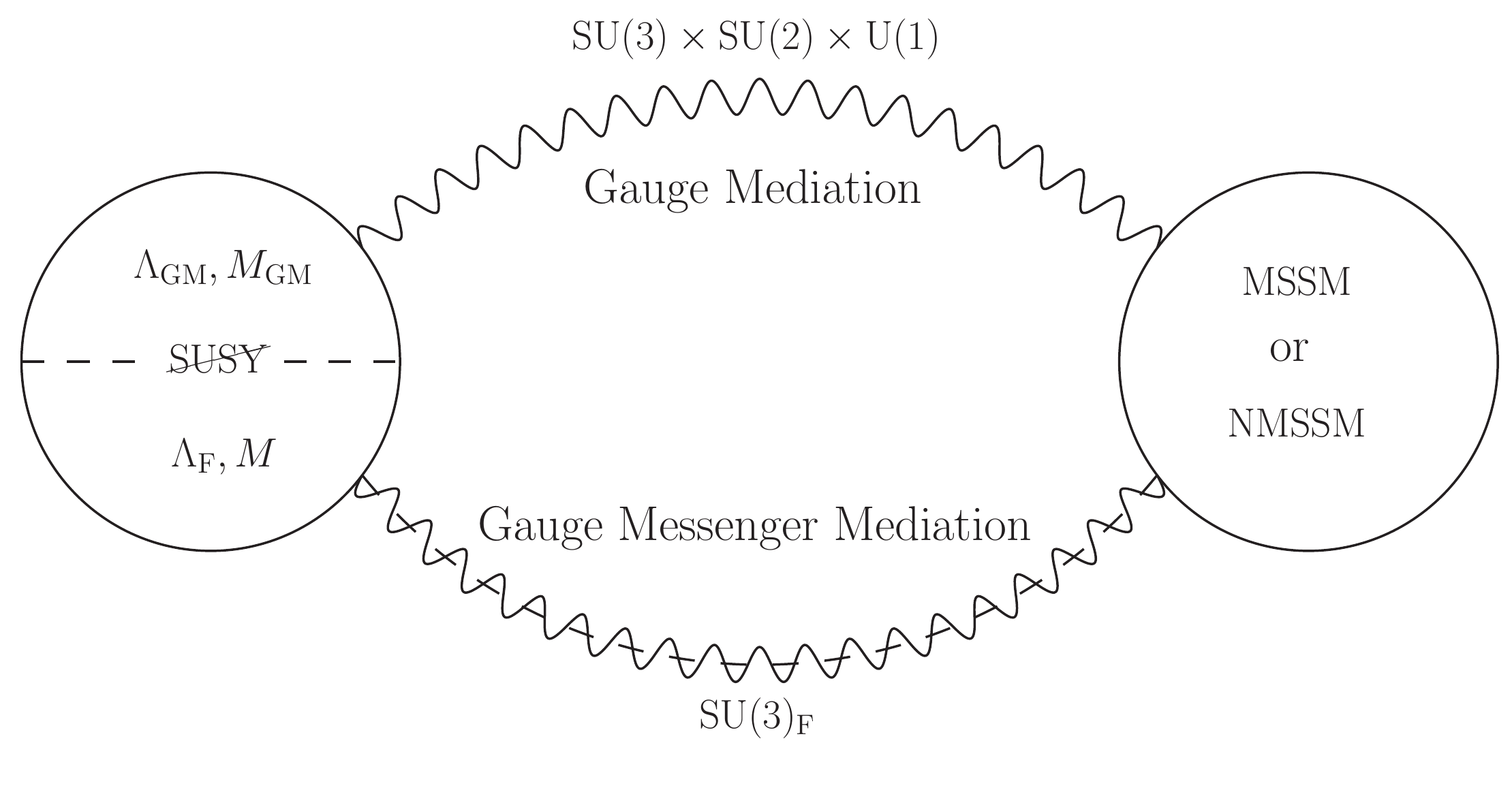}
  \caption{A  representation of gauge mediation of $SU(3)\times SU(2)\times U(1)$ combined with flavour-gauge messengers in $SU(3)_F$. This paper extends the framework by giving a dynamical explanation of $SU(3)_F$ SUSY breaking and the generation of the leading Yukawa structure in terms of magnetic SQCD.  }\label{figureofsetup}
\end{figure}

In detail, the models we will propose have the following properties:
\begin{itemize}

\item[$\clubsuit$] An ISS sector, with a weakly gauged $SU(3)_L\times  SU(3)_R\rightarrow SU(3)_V\equiv SU(3)_F$ subgroup of the full flavour symmetry. The fields that obtain $F$-term vevs contribute to the mass of the flavour gauge vector multiplet, which play the role of gauge messengers.

\item[$\clubsuit$] Fields charged under the flavour symmetry that play the role of flavons (these may or may not be part of the ISS sector itself).

\item[$\clubsuit$] The flavon and F-term vevs and hence the $SU(3)_F$ breaking all have a dynamical origin driven by the ISS sector.

\item[$\clubsuit$]  Flavon couplings to matter fields to generate Yukawa couplings.
\end{itemize}

With this general structure, a hierarchy can emerge between the $F$-terms and the one-loop induced vevs of many of the classical moduli including the flavons and other magnetic meson operators; consequently one can find an explanation of the Yukawa hierarchies that is closely aligned with a source of flavour mediation of SUSY breaking. 
The  effects from the flavour gauge mediation then combine with the standard gauge mediation of the SSM gauge groups $SU(3)_c\times SU(2)_L \times U(1)_Y$, yielding phenomenology with a number of interesting features:
\begin{itemize}
\item[$\spadesuit$]   One loop tachyonic soft mass terms from flavour gauge mediation that compete with the usual positive gauge mediated contribution.
\item[$\spadesuit$]  Typically, this leads to net tachyonic UV boundary conditions for stops, and hence lighter 3rd generation squarks at the weak scale, with improved naturalness properties.
\item[$\spadesuit$]  Mild flavour changing neutral currents from subleading off-diagonal mass terms, that decouple as $\alpha_F\rightarrow 0$.
\item[$\spadesuit$] For a weakly gauged system, $\text{tr}\mathcal{M}^2_{\text{Vector}}=-\text{tr}\mathcal{M}^2_{\text{Higgs}-\Phi}$ (i.e. the $\text{Str}{M^2}=0$ relation still holds).
\end{itemize}
The second entry in the above list is particularly important: although the stops may be tachyonic in the UV, the RGE contribution from  the 
gluino mass $M_3$ yields a net positive stop mass-squared at the weak scale. In terms of stability (or lack thereof) this indicates the possibility of charge and colour breaking minima along $F$- and $D$-flat directions. However such minima can be avoided\footnote{One might argue that morally speaking they are in any case no different from the global supersymmetric minimum of ISS and decay to them is likewise negligible.}. For example the $U_3D_3D_2$ direction has a potential of the form $V(\phi)\sim \phi^2 (m_{\tilde {t}_R}^2 + m_{\tilde {b}_R}^2+m_{\tilde {s}_R}^2)$. As long as there are enough positive mass-squareds in the UV this need never become unstable. Note that there is relative freedom in the choice of parameters as the gluino mass itself arises by the usual gauge mediation. The nett stop mass $m_{\tilde{t}}^2$ is therefore quite independent, and can be much lower than the gluino mass.  This is an improvement on usual gauge mediation (with respect to naturalness) as current bounds on gluinos (i.e. $1600$ GeV \cite{ATLAS-CONF-2013-047,Chatrchyan:2012lia}),   fix a supersymmetry breaking scale  
which would usually force stops to be at least this heavy.
Note that the Higgs soft terms are unaffected by flavour mediation as the Higgs doublets are $SU(3)_F$ singlets.  Indeed the Higgs soft terms are mediated by $SU(2)_L$ and $U(1)_Y$ and set at the messenger scale.  If one chooses the leptons to be charged under $SU(3)_F$ then the splitting extends to the sleptons as well, allowing for a light Stau NLSP.  Such a setup is consistent and compatible with supersymmetric Grand Unified Theories and indeed (if one wants to avoid additional SM charged states) mixed $SU(3)_F\times G_{SM}$ anomaly cancellation strongly suggests it.  

We should add that $SU(3)_F$ symmetries have been considered in the past as a good starting point for flavour structure \cite{Abel:2000hn,Abel:2001ur,King:2001uz,King:2003rf,Ross:2004qn,Antusch:2008jf}. Usually one requires a quasi-degeneracy between the first and second generation to avoid flavour changing effects. This may arise from an approximate remaining $SU(2)_F$ and we build this into some of our examples. In other examples this degeneracy will be broken mildly and therefore the discovery of (small) FCNC processes is a prediction of our framework.  Indeed an important advantage of combined flavour+standard gauge-mediation is that, with the first and second generation masses in the TeV range and dominated by the standard mediation contribution, the non-degeneracy required in the favour-mediation contribution is significantly relaxed, to the extent that it can be consistent to have no apparent $SU(2)_F$ symmetry for the first two generations, but just the correct general pattern of flavour dependence.

The outline of the paper is as follows. In section \ref{Dynamicaltop}  we demonstrate a simple example in which the top coupling is generated through a meson vev and supersymmetry breaking in the $SU(3)_F$ is generated dynamically. In section \ref{sectioncalcofsoftmasses} we give a derivation of the leading effect of the soft terms and discuss its limitations. We extend this in section \ref{magneticquarks} in which magnetic quark vevs generate a prototype Yukawa matrix.  In section \ref{perturbativemodelofflavour} we explore an entirely perturbative model of flavour in which the flavons are adjoint chiral superfields and the flavour breaking scale is generated dynamically.  In section \ref{susybreaking} we emphasise some of the features of the model's SUSY spectrum.  In section \ref{flavourconstraints} we look at possible flavour contraints on the model and ideas to address the strong CP problem as well as anomalies. Finally in section \ref{Discussion} we discuss how such a programme may be refined and we conclude.


\section{SUSY breaking \& flavour alignment with mesonic flavons}\label{Dynamicaltop}

We start with a simple case in which $SU(3)_F$ is broken to $SU(2)_F$ dynamically, and flavour and SUSY breaking ($F$-terms) are aligned. As described in the Introduction, we base the flavour/SUSY breaking sector on the magnetic description of Seiberg duality \cite{Seiberg:1994pq,Intriligator:1995au} with $N_f>3$ flavours.  For simplicity we assume a degenerate dynamical scale for the magnetic and electric theories, denoted by $\Lambda$. The additional 
$N_f-3$ flavours are necessary for the system to be in the metastable phase which requires $N_c+1\leq N_f<3N_c/2$. The obvious choice of choice $N_f=3$ is not within this window, and it may or may not result  in a metastable minimum since it would require $N_c=2$ and hence has $N_f=3N_c/3$. This is a marginal case and establishing whether or not it has metastable minima requires a two-loop treatment\footnote{Certainly, since the non-perturbative potential has no dependence on the dynamical scale $\Lambda$, the decay rate would not be parametrically suppressed. In addition we assume that the dynamical scale of the flavour sector is too large to disturb the metastability  
\cite{Forste:2006zc}.}.  Therefore we will for the moment consider general $N_f$, with the generic magnetic description being an $SU(\tilde{N}=N_f-N_c)$ gauge theory. 

Supersymmetry breaking is induced  as in the prototypical ISS model \cite{Intriligator:2006dd} and related works \cite{Kitano:2006xg,Abel:2007jx,Giveon:2007ef,Xu:2007az,Zur:2008zg,Abel:2008tx,Essig:2008kz,Koschade:2009qu}. We weakly gauge the $SU(3)_F$ flavour subgroup of the diagonal $SU(N_f)$ flavour symmetry before it is broken by the rank condition, to $SU(2)_F$. The flavour gauge fields will then act as messengers, generating suitably aligned tachyonic scalar soft terms. In this example we will take the mesons of the magnetic description of SQCD to act as the flavon fields. 

In order to reduce clutter, here and throughout we will focus on the states in the magnetic theory transforming in the $SU(3)_L\times SU(3)_R$ block of the flavour group. For the most part the remaining fields simply provide the necessary degrees of freedom to be in the metastable phase. We shall use a hat to denote generic states. Thus the generic $N_f\times N_f$ magnetic meson states are $\hat{\Phi}$, while the components in the $SU(3)_L\times SU(3)_R$ block are denoted ${\Phi}$. In the present example we shall take the latter to be the flavon fields. 

\renewcommand{\arraystretch}{1.2}
\begin{table}
\begin{center} 
\begin{tabular}{|c|c|c|} 
\hline \hline 
Field & $G_{SM}$ & $SU(3)_L\times SU(3)_R\rightarrow SU(3)_F$ \ \\
\hline 
\(\hat{Q}^{f}\) & \(({\bf 2}, \frac{1}{6},  {\bf 3}) \)   & $ ({\bf \overline{3}},1) $\\ 
\(\hat{L}^{f}\) &  \(({\bf 2},-\frac{1}{2}, {\bf 1}) \)   & $ ({\bf  \overline{3} },{\bf 1}) $ \\ 
\(\hat{H}_d\) &  \(({\bf 2},-\frac{1}{2}, {\bf 1}) \)   &  ({\bf 1},{\bf 1})  \\ 
\(\hat{H}_u\) &  \(({\bf 2}, \frac{1}{2}, {\bf 1}) \)   & ({\bf 1},{\bf 1})  \\ 
\(\hat{D}^{f}\) & \(({\bf 1}, \frac{1}{3}, {\bf \overline{3}}) \)& ({\bf 1},{\bf 3})  \\ 
\(\hat{U}^{f}\) & \(({\bf 1},-\frac{2}{3}, {\bf \overline{3}}) \) & ({\bf 1},{\bf 3}) \\
\(\hat{E}^{f}\) & \(({\bf 1}, 1, {\bf 1}) \) & $({\bf 1},{\bf 3 })$  \\
\hline
\(\hat{\nu}^{f}\) & \(({\bf 0}, 1, {\bf 1}) \) & $({\bf 1},{\bf 3 })$  \\
\hline \hline
\end{tabular} \caption{ Matter content of the $SU(3)_F$ gauged flavour model based on the MSSM. By inclusion of the right-handed neutrinos, all anomalies including mixed $G_{SM}$-$SU(3)_F$ anomalies are cancelled by this matter content.}\label{tableMSSM} \label{tableseesaw1}
\end{center} 
\end{table}
\renewcommand{\arraystretch}{}


\renewcommand{\arraystretch}{1.2}
\begin{table}
\begin{center} 
\begin{tabular}{|c|c|c|} 
\hline \hline 
Field & $SU(\tilde{N})_{\text{mag}}$ & $SU(3)_L\times SU(3)_R$ \ \\
\hline 
\(\Phi\) &1   & $ ({\bf 3,\overline{3}}) $\\ 
\(\varphi\) & ${\bf \square }$  &$ ({\bf \overline{3}},1) $  \\ 
\(\tilde{\varphi}\) &  $ {\bf \overline{ \square} } $ &  ({\bf 1},{\bf 3})  \\ 
\hline \hline
\end{tabular} \caption{The $SU(3)_L\times SU(3)_R$ block of the magnetic (macroscopic) description of the meson-as-flavon model.  $\tilde{N}=N_f-N_c$.  The global $SU(3)_L\times SU(3)_R$ symmetry is explicitly broken to the diagonal due to a $\delta W_{\text{mag}}\supset  \mu^2 \text{Tr}\Phi $ term.  The diagonal $SU(3)_F$ flavour group is then associated with an anomaly free SSM flavour symmetry.} \label{tablemacro}
\end{center} 
\end{table}
\renewcommand{\arraystretch}{}
\renewcommand{\arraystretch}{1.2}
\begin{table}
\begin{center} 
\begin{tabular}{|c|c|c|} 
\hline \hline 
Field & $SU(N_c)$ & $SU(3)_L\times SU(3)_R$ \ \\
\hline 
\(q\) & ${\bf \square }$  &$ ({\bf \overline{3}},1) $  \\ 
\(\tilde{q}\) &  $ {\bf \overline{ \square} } $ &  ({\bf 1},{\bf 3})  \\ 
\hline \hline
\end{tabular} \caption{The $SU(3)_F$ charged fields of the Electric (microscopic) description of the flavon model. The electric quark mass $W= \text{Tr}(\hat{m} \hat{q}\hat{\tilde{q}})$ breaks the $SU(3)_L\times SU(3)_R$ flavour symmetries to the diagonal.} \label{tablemicro}
\end{center} 
\end{table}
\renewcommand{\arraystretch}{}

The matter content of the magnetic description is represented in table \ref{tablemicro}. As usual the superpotential is
\be
W_{\text{mag}}=h\text{Tr}\hat{\varphi}\hat{ \Phi}\hat{ \tilde{\varphi}}-h \text{Tr}({\hat{\mu}}^2\hat{\Phi}).
\ee
The second term (the operator which is dual to a quark mass-term in the electric theory) preserves the vector part of the $SU(3)_L\times SU(3)_R$ flavour symmetry that 
we are gauging, but otherwise is generic. By flavour rotations it can be diagonalised as
\begin{equation} 
\hat{\mu}^2 = \left( 
\begin{array}{cc} 
\mu^2 I_{3\times 3} & 0 \nonumber \\
0 & \tilde{\mu}^2  \end{array} \right)
\end{equation}
where $\tilde{\mu}^2 $ is a diagonal $(N_f-3)\times (N_f-3)$ matrix, with generic entries. They could for this entire discussion be zero, but in order to break flavour correctly we must assume that they obey $\tilde{\mu}^2_{ii} < \mu^2$ $\forall i$.

Due to the ISS rank condition\footnote{The matrix $(\varphi \tilde{\varphi} )_{ij}$ can only have maximum rank $\tilde{N}$ or $N_f$ whichever is smaller.} supersymmetry is broken when $N_f>\tilde{N}$. Under our assumption that $\mu^2 $ are the largest entries in the $\hat{\mu}^2$ matrix, $\tilde{N}$ of the $SU(3)_F$ flavours get vevs becoming heavy along with the off-diagonal ${SU(3)_L\times SU(N_f-3)_R}$ and ${SU(N_f-3)_L\times SU(3)_R}$ components of $\hat{\Phi}$. 
It is worth displaying the remaining relevant $SU(3)_L \times SU(3)_R$ block of the tree-level scalar potential in detail: 
\be
V=\sum_{i,j}|(\varphi \tilde{\varphi})_{ij}-\mu^2\delta_{ij}|^2+ \sum_{i,\alpha}|(\varphi \Phi)_{i,\alpha}|^2+ \sum_{i,\alpha}|( \Phi \tilde{\varphi})_{\alpha,i}|^2,
\ee
where $i,j$ are flavour indices and $\alpha,\beta$ are $SU(\tilde{N})$ indices.  

As the ISS breaking should also break $SU(3)_F$ spontaneously to $SU(2)_F$, we have an additional requirement that $\tilde{N}<3$, or $N_f<N_c+3$, on top of the requirement that $N_c+1 \leq N _f < 3N_c/2$. Therefore the minimal case would have $N_c=3$ and $N_f=4$ (with $\tilde{N}=1$) and would actually be $s$-confining with no magnetic gauge group. The case $N_c=4$ allows $N_f=5$ (also $s$-confining), and $N_c=5$ allows $N_f=6$ ($s$-confining)  or $7$ ($SU(\tilde{N}=2)$ magnetic gauge group). Higher $N_c$ obviously allow $\tilde{N}=3$ as well, although because this would be universal spontaneous breaking of the gauged $SU(3)_F$ symmetry it is not interesting for the present discussion. 

Thus $\tilde{N}=1$ or $2$ flavours of quark in the $SU(3)_F$ block  get vevs, and the $SU(3)_F$ symmetry is broken spontaneously. Meanwhile the extra $N_f-3$ flavours remain vevless and indeed the $SU(N_f-3)_V$ symmetry is broken only by the explicit $\tilde{\mu}^2_{ii}$ non-degeneracy in this model. 

Let us take $\tilde{N}=2$ as an example. The vevs of the $SU(3)_F$ flavoured quarks may be written as 
\be
\varphi^T=\tilde{\varphi}=\left( \begin{array}{c} \mu \\ \mu  \\ 0 \end{array}\right)\, .
\ee
Since $SU(3)_F$ is weakly gauged, the spontaneous breaking provides the goldstone modes to generate massive flavour gauge bosons, by the supersymmetric Higgs effect.  The minimum of the potential still contains an $SU(2)$ symmetry.  The $F$-terms in this block are given by 
\be
F_{\Phi}=\left( \begin{array}{ccc} 0&0 & 0 \\0 &0 &0 \\ 0&0 &h\mu^2 \end{array}\right)
  \ \ \   \text{such that}  \ \ \ \ V_{\text{min}}=|h^2\mu^4|.
\ee
Note that there are no other $F$-terms in the $SU(N_f-3)_L\times SU(N_f-3)_R$ block. 

We find that the flavour mediated contribution to the soft terms due to the gauging of $SU(3)_F$ is 
(see section \ref{sectioncalcofsoftmasses})
\be
\delta m^2_{Q,U,D}=-\frac{g^2_F}{16\pi^2} {|h^2\mu^2|}\left(
\begin{array}{ccc}
 \frac{8}{9 } & 0 & 0 \\
 0 & \frac{8 }{9} & 0 \\
 0 & 0 & \frac{20}{9} \\
\end{array}
\right)\label{eq:softmodel1}+...
\ee
plus subleading off-diagonal terms. Note the minus sign: this contribution clearly suppresses third generation scalars relative to 1st and 2nd (see figure \ref{fig:ModelA}) and when combined with usual gauge mediation already achieves our first aim of dynamically generating a split squark spectrum suitable for a natural model of supersymmetry breaking.

\subsection{Flavour from magnetic mesons} 
Of course to complete the programme we need a flavour model. As the mesons are composite objects the model of flavour inevitably has to have a strong coupling element to it.  In models such as this one where the goldstino partly resides in the flavon, the main additional requirement is a vev for the scalar component, which unavoidably breaks the (accidental) $R$-symmetry of the model spontaneously. Breaking $R$-symmetry was the subject of a number of studies in the past, due to its central role in the generation of a Majorana gaugino mass. Here we seek a suitable deformation of the theory that can achieve this without explicitly breaking the $SU(3)_F$ gauge symmetry of the Lagrangian. 
We choose \cite{Giveon:2007ef,Xu:2007az,Essig:2008kz,Koschade:2009qu}\footnote{Note that we cannot for example use the operator $\delta W\supset  h^2 m_Y\text{Tr}(\tilde{Y}Y)$ of ref.\cite{Kitano:2006xg}   which generates $\braket{X}=\frac{1}{2}hm_Y$, because it breaks the $SU(3)_F$ explicitly.}  
\be
\delta W= \frac{1}{2}h^2\mu_{\Phi}(\text{Tr}\Phi^2 +\gamma\text{Tr}(\Phi)^2).
\ee
It is then expected that the one loop Coleman-Weinberg potential
 will minimise the classical modulus as
\be
{\Phi}=\left( \begin{array}{ccc} 0&0 & 0 \\0 &0 &0 \\ 0&0 & \braket{X} \end{array}\right),
\ee
so that the non-zero top Yukawa is aligned with the lightest squark generation as required. 
Indeed considering the charges of the fields in tables \ref{tableMSSM} and \ref{tablemacro}, 
and matching the magnetic  meson with its electric composite $\Phi \Lambda = q\tilde{q}$,
the allowed Yukawas are
\be
W_{\rm Yuk}={\lambda_u}\frac{\Lambda}{\Lambda_0^2} H_u Q \Phi U + {\lambda_d} \frac{\Lambda}{\Lambda_0^2} H_d Q \Phi D \label{mesoncase},
\ee
where $\Lambda_0$ is some physical scale in the electric theory while recall that $\Lambda$ is identified with the dynamical transmutation scale of the ISS theory.
 
 We can make a completely general observation that will apply to all the models of strongly coupled flavour that we consider, which arises from the fact that 
 the Yukawa coupling for the top is $\lambda_u \braket{X}\Lambda/\Lambda_0^2 \lesssim \mu \Lambda/\Lambda_0^2 $. Since phenomenology requires this 
 to be ${\cal O}(1)$ we require $\Lambda_0\ll \Lambda$. In other words the operators present in $W_{\rm Yuk}$ creating the Yukawa couplings are generated by integrating out 
 degrees of freedom at a relatively light scale $\Lambda_0$ that lies somewhere between $\Lambda$ and $\mu$. These degrees of freedom must mix with the composite states but it is easy to see that they can be charged only under flavour and hence are simply spectators of the ISS sector, so they need not interfere with the metastability there. When we come to consider weakly coupled flavour models later we shall consider the complete picture.
  
 The model above not only gives the spectrum of natural supersymmetry but also generates  the top coupling dynamically; indeed dynamical alignment of flavour with 
 SUSY breaking demands it. Unfortunately it does not naturally yield further couplings.  This is because the only other fields with vevs are 
 the quarks, but since $\braket{X}\lesssim \mu$ is loop suppressed, operators involving them are likely to be larger than the top coupling itself. We would actually have to suppress all couplings between SSM matter and the magnetic quarks if we were to pursue this as a model of flavour.  
 In the next sections we will explore examples that  give a more realistic flavour model.

\section{Calculation of the soft masses}\label{sectioncalcofsoftmasses}
Before continuing, we briefly recall the calculation of the soft mass contribution from flavour gauge messengers, in the flavour basis -- this was the result that was used in the previous section. The sfermion soft masses of the visible sector, at one loop, are given by \cite{Intriligator:2010be,Brummer:2013upa}
\be
\delta m_{Q_I}^2 = - \mathbf{T}^{ab}_{IJ} \delta_{IJ}\mathcal{Z}^{ab}|_{\theta^4} \ \ , \ \ \mathbf{T}^{ab}=\{ T^a,T^b\}.
\ee
where a,b are generator indices running over $1,...,\text{dim} G$ and $I$ is the sum over all charged Chiral superfields.  This term arises from the one-loop K\"ahler potential 
\be
K_{eff}^{(1)}=\sum_{I,J}Q^{\dagger}_{I}\mathbf{T}^{ab}_{IJ}Q_{J}\delta_{IJ} \mathcal{Z}^{ab}+...
\ee 
where the $Q_I$'s are visible sector Chiral superfields and $\Phi_i$ will denote the hidden sector fields. The  matrix $\mathcal{Z}^{ab}$ is defined as 
\be
\mathcal{Z}^{ab}= \left(\frac{\alpha_F}{4\pi}\right)\text{log}\left(g^2_F\sum_{i}  \frac{\Phi_i^{\dagger} \mathbf{T}^{ab}_i\Phi_i}{\Lambda^2}\right)  .\label{masterformula}
\ee
The combination $\Phi_i^{\dagger} \mathbf{T}^{ab}_i\Phi_i$ is traced over flavour indices where $i$ runs over each field: it is a set of matrices in the $\theta$-expansion that terminate at $\theta^4$ with  $(\bar{F},F)^{ab}$.  The inner product is defined as 
\be
(\bar{\phi},\phi)^{ab}\equiv \sum_{i}\phi^{\dagger}_i\{ T^a,T^b \}\phi_i.
\ee
 Typically (and in our case) these matrices are \emph{not} diagonal and one cannot assume $(\bar{\phi},\phi)^{ab}=(\bar{\phi},\phi)\delta^{a'b'}$.  The result is that whilst one may expand the logarithm as 
\be
\text{Log} [(\bar{\phi},\phi) +\theta^2 (\bar{\phi},F) + \bar{\theta}^2(\bar{F},\phi) +\theta^2\bar{\theta}^2(\bar{F},F)  ]^{ab}
\ee
one cannot take the $\text{Log}[A^{ab} (\mathbf{I}^{ab}+B^{ab}/A^{ab}) ]$ and carry out an expansion in $F/M$.  One must instead take the the matrix Logarithm of \refe{masterformula} and give an approximate expansion, typically in the off-diagonal elements of $(\bar{\phi},\phi)$. If one can carry out the expansion, then the soft mass matrix can be represented as 
\be
(m^{2(1)}_I)_{cd}=- \left(\frac{\alpha_F}{4\pi}\right) \sum_{a,b} \mathbf{T}^{ab}_{cd}\left(\frac{(\bar{F},F)^{ab}(\bar{\phi},\phi)^{ab} -  (\bar{F},\phi)^{ab}(\bar{\phi},F)^{ab}  }{(\bar{\phi},\phi)^{ab}(\bar{\phi},\phi)^{ab}  }\right)+O(|F|^4).
\ee
The result above is found from the usual small $F/M$ expansion and generalising $X=M+\theta^2 F$, where   $\alpha_F=g^2_F/4\pi$ is the $SU(3)_F$ fine-structure constant.  This will typically give the leading contribution, with subleading off-diagonal entries, and so we can use this formula to describe the essential feature of the model. More details may be found in \cite{Intriligator:2010be}.

\section{The magnetic quark case}\label{magneticquarks}
As we have seen, typically the magnetic quark vevs are larger than the one loop induced meson vev, so a natural way to generate Yukawa structure in a strongly coupled theory of flavour is to begin with the magnetic quark vevs. In this section we give two examples of dynamical supersymmetry breaking in which the magnetic quark vevs are used to generate a prototypical Yukawa structure.

\subsection{$\tilde{N}=1$ example}
The case $\tilde{N}=1$ 
allows for the vevs 
\be
\varphi^T=\tilde{\varphi}=\left( \begin{array}{c} 0 \\0  \\ \mu \end{array}\right)
\ee
which breaks  $SU(3)_F\rightarrow SU(2)$. The leading Yukawa terms can be inferred by baryon matching. In this case (assuming for simplicity 
equal dynamical scales in the electric and magnetic theories) we have 
\begin{equation}
\Lambda^{\tilde{N}} q^{N_c} \sim \Lambda^{N_c} \varphi^{\tilde{N}}\, ,
\end{equation}
so when $\tilde{N}=1$ the natural object appearing in the superpotential is $ \left( \frac{\Lambda}{\Lambda_0}\right)^{N_c} \varphi/\Lambda $ and similar for $\tilde{\varphi}$. 
Taking the minimum case of $N_c=3$ would give a leading contribution to the up-Yukawas of the form  
\be
W\supset  \frac{\Lambda ^4}{\Lambda_0^6}{(\varphi U) ( \tilde{\varphi}Q)}H_u\label{eq:Yuk}\, .
\ee
Now an order one top-Yukawa requires $\mu \Lambda^2/\Lambda_0^3\sim {\cal O}(1)$, but our previous comments apply: we require $\mu < \Lambda_0 < \Lambda$ for the 
picture to be valid. 

The classical moduli space has maximal symmetry when $\braket{\Phi}=0$ and the supersymmetry breaking is therefore \emph{anti-aligned} with flavour:
\be
F_{\Phi}=\left( \begin{array}{ccc} h\mu^2&0 & 0 \\0 &h\mu^2&0 \\ 0&0 &0 \end{array}\right)
  \ \ \   \text{such that}  \ \ \ \ V_{\text{min}}=2 |h^2\mu^4|.
\ee
The soft mass formula is 
\be
\delta m^2_{Q,U,D}=-\frac{g^2_F}{16\pi^2} {|h\mu|^2}\left(
\begin{array}{ccc}
 \frac{11}{18 } & 0 & 0 \\
 0 & \frac{14}{18} & 0 \\
 0 & 0 & \frac{29}{18} \\
\end{array}
\right)+...\label{eq:softmassesmodel2}
\ee
Thus this model also generates a mild splitting between 1st and 2nd generation squarks. An example low-energy spectrum is shown in 
figure \ref{fig:ModelB}.

\subsection{Flavour in the $\tilde{N}=1$ model}
As we saw above the third generation is anti-aligned with supersymmetry breaking. In this model the magnetic meson may contribute subleading effects.  The Coleman-Weinberg potential generates vevs for the diagonal components of the magnetic meson
of the form 
\be
{\Phi}=\left( \begin{array}{ccc} \braket{x}&0 & 0 \\0 &\braket{x} &0 \\ 0&0 & \braket{y}\end{array}\right).
\ee
As before $\braket{x}$ may be generated by additional mesonic operators in the superpotential, and it is typically suppressed with respect to the leading term. The term $\braket{y}$ may also arise due to additional operators of magnetic quarks in the superpotential and is typically smaller than $\braket{x}$. Defining 
\be
\epsilon_x=\frac{\braket{x}\Lambda }{\Lambda_0^2} \ \ \epsilon_y=\frac{\braket{y} \Lambda}{\Lambda_0^2}\, ,
\ee
\refe{mesoncase} leads to a contribution to the up-Yukawas of 
\be
\delta Y_u\sim \left( \begin{array}{ccc} \epsilon_x&0 & 0 \\0 &\epsilon_x &0 \\ 0&0 &\epsilon_y\end{array}\right)\label{eq:meson2}\, .
\ee
All scales have been generated dynamically either from the vev of an electric quark mass or a one-loop Coleman-Weinberg minimisation of the scalar potential. However the model does not yet break the residual $SU(2)_F$ flavour symmetry. In addition the up and down Yukawas are still diagonal in a common basis. The former problem is quite difficult to solve in the $\tilde{N}=1$ model. Moreover off-diagonal components can also not easily be generated. For example one might consider additional operators involving the magnetic quarks. Assuming that $U$ is in a fundamental triplet of flavour, while $Q$ is in an antifundamental (in order to for example avoid mixed $SU(3)_F$ Standard Model anomalies), then $SU(3)_F$ baryonic operators such as  
\be
W\supset \hat{O}_b=\frac{\Lambda^6}{\Lambda_0^9} \tilde{\varphi} U  (\varepsilon \tilde{\varphi} \tilde{\varphi} Q)H_u \label{ifBaryon}\nonumber
\ee
are bound to give zero contribution to the Yukawa couplings since $\varphi$ and $\tilde{\varphi}$ have only one non-zero component. To summarise the $\tilde{N}=1$ model, the operator \refe{eq:Yuk} nicely generates a natural Yukawa coupling, and \refe{eq:meson2} may contribute some subleading effects but leaves the flavour structure diagonal and still $SU(2)_F$ symmetric. 
\subsection{The $\tilde{N}=2$ example}
These problems may be partially solved if one also includes an $\tilde{N}=2$ model. First consider the $\tilde{N}=2$ model in isolation. For $SU(\tilde{N})=SU(2)$ the rank condition gives (recalling that this is for just the $SU(3)_F$ charged degrees of freedom)
\be
\varphi= e^{-i\theta}\left( \begin{array}{c} 0 \\ {\mu}\\ {\mu} \end{array}\right) \ \ ,  \ \ \tilde{\varphi}^T=e^{i\theta}\left( \begin{array}{c} 0\\ {\mu} \\ {\mu}  \end{array}\right) \, ,
\ee
spontaneously breaking $SU(3)_F\rightarrow U(1)$, where we have assumed a relative phase between the magnetic quark vevs.  

As we saw earlier the minimal case has $N_c=5$ so that the natural scaling in the 
superpotential is $ \left( \frac{\Lambda}{\Lambda_0}\right)^{5/2} \varphi/\Lambda $, and 
the Yukawa operator is now 
\be
W\supset  \frac{\Lambda ^3}{\Lambda_0^5}{(\varphi U) ( \tilde{\varphi}Q)}H_u\label{eq:Yuk22}\, ,
\ee
which gives
\be
Y_u\sim \frac{\Lambda ^3 \mu^2}{\Lambda_0^5} \left( \begin{array}{ccc}0&0 & 0 \\0 &1  &1 \\ 0& 1&1 \end{array}\right) \, .   
\ee
The Coleman-Weinberg potential will set
\be
{\Phi}=\left( \begin{array}{ccc} \braket{x}&0 & 0 \\0 &\braket{y} &0 \\ 0&0 & \braket{y}\end{array}\right)
\ \ \ ,  \ \ 
F_{\Phi}=\left( \begin{array}{ccc} h\mu^2&0 & 0 \\0 &0 &0 \\ 0&0 & 0 \end{array}\right)
  \ \ \   \text{such that}  \ \ \ \ V_{\text{min}}=|h^2\mu^4|
\ee
and the soft masses are now found to be
\be
\delta m^2_{Q,U,D}=-\frac{g^2_F}{16\pi^2} {|h\mu|^2}
\left(
\begin{array}{ccc}
 \frac{48+\sqrt{29}}{45 } & 0 & 0 \\
 0 & \frac{42-2\sqrt{29} }{45} & 0 \\
 0 & 0 & \frac{66+\sqrt{29}}{45} \\
\end{array}
\right)+...
\ee
An example spectrum is shown in figure \ref{fig:ModelC}.  The soft terms do not preserve an $SU(2)$ at leading order and so it would likely lead to larger FCNCs than the previous two models. As we later discuss in more detail, to some extent this is alleviated by reducing the value of $g_F$ but at the cost of lessening the effect of reduced stop masses and therefore decreasing the naturalness effect, as pictured in figure \ref{fig:ModelC}.  Thus there is an interesting interplay between FCNCs and naturalness in this model.

\subsection{Two ISS sectors}
 A more complete model of flavour is then to include two ISS sectors: the first $\tilde{N}=1$ component (referred to as ISS(1)) generates the leading soft breaking of \refe{eq:softmassesmodel2} and top Yukawa of \refe{eq:Yuk}, while the second $\tilde{N}=2$ component (referred to as ISS(2))  generates subleading Yukawas.  The subleading soft-terms  from ISS(2) can be ignored and the resulting spectrum is represented in figure \ref{fig:ModelB}.  
 
 The magnetic quarks of ISS(2) $\varphi_2,\tilde{\varphi}_2$ give operators
\be
W\supset \mathcal{O}_2=\frac{\lambda^u_2\Lambda_2^3 }{\Lambda_0^5}(\varphi_{2}U)(\tilde{\varphi_2}Q)H_u\, .
\ee
In addition one can incorporate all other operators including those coupling ISS(1) to ISS(2), such as (for example) 
$
(\tilde{\varphi}_i\varphi_j), (\tilde{\varphi}_i Q), (\varphi_i U ),
(\varphi_1\varphi_2 Q)$ and    $(\tilde{\varphi}_1\tilde{\varphi}_2 U)$. Thus one has operators 
\be 
\mathcal{O}_3 \sim (\tilde{\varphi}_{i}\varphi_j)(\tilde{\varphi}_{1}Q)(\varphi_{1}U) H_u
\ee
\be 
\mathcal{O}_4\sim (\tilde{\varphi}_{i}\varphi_j)(\tilde{\varphi}_{1}Q)(\varphi_{2}U)H_u
\ee

\be 
\mathcal{O}_5\sim (\tilde{\varphi}_{i}\varphi_j)(\tilde{\varphi}_{2}Q)(\varphi_{1}U)H_u
\ee
\be 
\mathcal{O}_6\sim (\tilde{\varphi}_{i}\varphi_j)(\tilde{\varphi}_{2}Q)(\varphi_{2}U)H_u
\ee
and further operators suppressed by powers of $\mu^2 /\Lambda_0^2$ such as 
\be
\mathcal{O}_7\sim (\tilde{\varphi}_i\varphi_j)  (\varphi_1\varphi_2 Q)   (\varphi_k U ) H_u
\ee
\be
\mathcal{O}_8\sim (\tilde{\varphi}_i\varphi_j) (\tilde{\varphi}_k Q )   (\tilde{\varphi}_1\tilde{\varphi}_2 U)  H_u\, .
\ee
Additional discrete symmetries may be applied here, judiciously, and clearly there is enough freedom here to generate realistic looking Yukawa matrices.

\section{Alignment with UV complete models: a dynamical flavour scale}\label{perturbativemodelofflavour}
As we mentioned earlier, the fact that for example the combination $\mu\Lambda/\Lambda_{0}^{2}$
appears in the Yukawa couplings, where $\Lambda_{0}$ is a mass scale
in the UV completion of the theory, while $\Lambda$ is the strong
coupling scale with $\mu<\Lambda$, inevitably means that there must
be additional physics below the strong coupling scale. It is therefore
of interest to present a model that is entirely perturbative apart
from any ISS sectors that we add to break gauged $SU(3)_{F}$ and
SUSY, that has no non-renormalizable operators, and in which all the
additional modes are explicit. 

\subsection{The perturbative flavour structure}
The idea we focus on here is that the flavons correspond to elementary
flavour adjoints $\phi_{u,d}$. As well as providing the necessary
flavour structure these can ameliorate or even eliminate the strong
CP problem \cite{Masiero:1998yi,Abel:2000hn,Abel:2001ur}. In addition we will require additional fields
of mass ${\cal O}(\Lambda_{0})$ charged under the Standard-Model
gauge group. These are denoted $D$ and $\tilde{D}$, and carry the same hypercharge
as the usual down multiplets and its conjugate respectively. (In order to have correct flavour structure
we would obviously repeat the following procedure for the ups.)

Let us begin this time with the generation of flavour structure, which
appears in the superpotential as 
\begin{equation}
W\supset qhD+\Lambda_{0}\tilde{D}^{T}D+\tilde{D}^{T}\phi_d  d,
\end{equation}
where $\Lambda_{0}$ is our dimensionful parameter, and as promised
all the operators are now renormalizable. For one generation, upon integrating out the
heavy $D,\tilde{D}$ we would have light and heavy quarks given by 
\begin{equation}
\left(\begin{array}{c}
d_{\ell}\\
d_{h}
\end{array}\right)=\left(\begin{array}{cc}
c & s\\
-s & c
\end{array}\right)\left(\begin{array}{c}
d\\
D
\end{array}\right)
\end{equation}
where (taking a classical background for $\phi_d )$ 
\begin{equation}
\tan\theta=-\phi_d /\Lambda_{0}.
\end{equation}
 More generally, with multiple generations, the heavy state is given by 
\begin{equation}
d_{h}=\frac{1}{\sqrt{\Lambda_{0}^{2}+\frac{1}{3}\mbox{Tr}(\phi_d ^{2})}}\left(\Lambda_{0}D+\phi_d  d\right)
\end{equation}
where we normalize as $d_{i}^{\dagger}d_{j}=\delta_{ij}/{3}$. The
orthogonal light state is then 
\begin{equation}
d_{\ell}=\frac{1}{\sqrt{\Lambda_{0}^{2}+\frac{1}{3}\mbox{Tr}(\phi_d ^{2})}}\left(-\phi_d  D+\Lambda_{0}d\right).
\end{equation}
Then the low-energy superpotential becomes 
\begin{equation}
W\supset qh\sqrt{\Lambda_{0}^{2}+\frac{1}{3}\mbox{Tr}(\phi_d ^{2})}(\Lambda_{0}^{2}-\phi_d ^{2})^{-1}\phi_d  d_{\ell}.
\end{equation}
Obviously in the limit that $\phi_d /\Lambda_{0}\ll1$ this can be
approximated as 
\begin{equation}
W\supset\frac{1}{\Lambda_{0}}qh\phi_d  d_{\ell},
\end{equation}
but interestingly one can have a natural expansion in flavour structure.

\subsection{The symmetry breaking}
We now wish to induce the required flavour symmetry breaking via the ISS mechanism to generate the flavon vev at the same time as SUSY is broken. It
is clear that we can never generate a satisfactory hierarchy with
a tracelsss adjoint. Therefore instead we consider $\phi_d $ to
have a trace component as well (i.e. there is an extra singlet). This
could have its own coupling but for simplicity we assume it is degenerate
with the other couplings so the full global symmetry (and possibly
even gauge symmetry) is $U(3)_{F}$. 

Consider two ISS theories
with $\tilde{N}=\tilde{N}'=1$. The magnetic theories
are the usual ISS theories with mesons $\Phi,\,\Phi'$, and we add
additional couplings between the elementary and mesonic adjoints.
Note that further complication will be required because we are attempting to incorporate the correct breaking of 
the $U(3)^5$ flavour symmetry, and multiple ISS sectors are a rather efficient way to do this dynamically.
However our aim is not to achieve a better model of flavour, but rather to focus on how the 
SUSY breaking can naturally be aligned with it. 

The $SU(3)_F$ charged part of the superpotential is 
\be
W_{ISS} =\left( \Phi\tilde{\varphi}\varphi-\mu^{2}\Phi+m\phi_d \Phi\right)
  +\left(\Phi'\tilde{\varphi}'\varphi'-\mu'^{2}\Phi'+m\phi_d \Phi'\right).
\ee
It contains the usual ISS terms, and in addition terms coupling the
elementary adjoints to the meson of the original SQCD theory. 
Given
that the latter correspond to renormalizable couplings in the electric
theory, one expects $m\sim\Lambda$ and $m'\sim\Lambda'$, while $\mu=m_Q \Lambda$ and $\mu'=m_{Q'}\Lambda'$ depend on 
the dimensionful mass parameters, so are not directly related.
Note that as per the previous models we can 
more or less ignore the $SU(N_f-3)_V$ part of the superpotential which simply provides extra quark supermultiplets, provided that 
their linear terms have $\tilde{\mu}^2_{ii} < \mu^2$. 

The potential for this part of the theory is of the following form
at tree-level; 
\begin{eqnarray}
V_{ISS} & = & \sum_{b=1}^{3}|\Phi_{ab}\tilde{\varphi}^{a}|^{2}+|\varphi^{a}\Phi_{ab}|^{2}+|\Phi'_{ab}\tilde{\varphi}^{\prime a}|^{2}+|\varphi^{\prime a}\Phi'_{ab}|^{2}\nonumber \\
 &  & +\sum_{ab}|\varphi_{a}\tilde{\varphi_{b}}-\mu^{2}\delta_{ab}+m\phi_{d,ab}|^{2}+|\varphi'_{a}\tilde{\varphi}'_{b}-\mu'^{2}\delta_{ab}+m'\phi_{d,ab}|^{2}\nonumber \\
 &  & +\sum_{ab}\left|m\Phi_{ab}+m'\Phi'_{ab}+\tilde{D}_{a}d_{b}\right|^{2}.
\end{eqnarray}
Since the same $\phi_d $ couples to both ISS sectors, it is not
possible to satisfy both unbroken SUSY constraints at the same time. In
order to see this we may write the $\phi_d $ VEV as 
\begin{equation}
\phi_d =\rho_{a}\delta_{ab}.
\end{equation}
Only one $(\tilde{\varphi}\varphi)_{ab}$ composite meson VEV is available
in each ISS sector for cancelling $F_{\Phi}$ terms. Hence we may
without loss of generality rotate to a basis in which $(\tilde{\varphi}\varphi)_{ab}=\eta^2_{1}\delta_{a1}\delta_{b1}$
and $(\tilde{\varphi}'\varphi')_{ab}=\eta^{\prime 2}_{3}\delta_{a3}\delta_{b3}$,
so that the ISS part of the potential is 
\begin{eqnarray}
V_{ISS} & = & (\eta_{1}^{2}-\mu^{2}+m\rho_{1})^{2}+(\mu^{2}-m\rho_{2})^{2}+(\mu^{2}-m\rho_{3})^{2}+ \nonumber\\
 &  & (\mu'^{2}-m'\rho_{1})^{2}+(\mu'^{2}-m'\rho_{2})^{2}+(\eta_{3}'^{2}-\mu'^{2}+m'\rho_{3})^{2} \nonumber \\
 & \implies & (\mu^{2}-m\rho_{2})^{2}+(\mu^{2}-m\rho_{3})^{2}+(\mu'^{2}-m'\rho_{1})^{2}+(\mu'^{2}-m'\rho_{2})^{2}.
\end{eqnarray}
The minimum is at
\be
\rho_{1}=\frac{\mu'^{2}}{m'}  \ \  , \ \ 
\rho_{3} = \frac{\mu^{2}}{m} \ \ , \ \ 
\rho_{2} = \frac{m^{2}\rho_{3}+m'^{2}\rho_{1}}{m'^{2}+m^{2}},
\ee 
with $F-$terms 
\begin{equation}
\frac{F_{\Phi_{i}}}{m'}=-\frac{F_{\Phi_{i}'}}{m}=\frac{mm'}{m'^{2}+m^{2}}(\rho_{1}-\rho_{3})\delta_{i2}.
\end{equation}
This gives 
\begin{equation}
\eta_1^2=m (\rho_3-\rho_1)\,\, ; \, 
\eta_3^{\prime 2}=m' (\rho_1-\rho_3).
\end{equation}
In order to generate a Yukawa hierarchy, suppose 
$\mu'=\xi^{3}\mu$ and $m'=m/\xi^{2}$
with $\xi\lesssim1$. Then $\rho_{1}=\xi^{8}\rho_{3}$ and $\rho_{2}\approx\xi^{4}\rho_{3}$.
Here the parameter $\xi\approx 0.4$ to achieve the correct down hierarchy. (The ups would require $\xi \rightarrow \xi^2 \approx 0.16$). Retaining the leading terms in $\xi^2$ only we find 
\be
\phi_d   =  \frac{\mu^2}{m} {\cal O} 
\left(\begin{array}{ccc}
\xi^{8} & 0 & 0\nonumber \\
0 & \xi^{4} & 0 \\
0 & 0 & 1 \end{array}\right)
\ee
leading to the correct Yukawa structure and for the supersymmetry breaking and gauge symmetry breaking,
\be
F_\Phi  = {\mu^2} {\cal O} 
\left(\begin{array}{ccc}
0 & 0 & 0\\
0 & 1 & 0 \\
0 & 0 & 0 \end{array}\right) \, ; \,
F_{\Phi'}  =  {\mu^2} {\cal O} 
\left(\begin{array}{ccc}
0 & 0 & 0 \\
0 & \xi^2 & 0 \\
0 & 0 & 0 \end{array}\right) \, 
\ee
\be
\varphi \tilde{\varphi}   =  {\mu^2} {\cal O} 
\left(\begin{array}{ccc}
1 & 0 & 0 \\
0 & 0 & 0 \\
0 & 0 & 0 \end{array}\right) \, ; \,
\varphi' \tilde{\varphi}'  =  {\mu^2} {\cal O} 
\left(\begin{array}{ccc}
0 & 0 & 0 \\
0 & \xi^2 & 0 \\
0 & 0 & 0 \end{array}\right) .
\ee
Clearly by adding further $\tilde{N}=1$ ISS sectors and adjoints $\phi_u$ for the ups, one can fill up the entire flavour structure. (This method of flavour breaking is 
direct in the sense that there is a field associated with each layer of symmetry breaking.) It should be noted however that only one set of susy breaking is required as it will be mediated to all fields charged under $SU(3)_F$. For example it could be that the up Yukawas are driven by an SQCD theory that has no metastability and does not contribute to supersymmetry breaking. Then the flavour mediated contribution to the soft-terms would all be aligned with the down Yukawas as described above.

For definiteness we consider this example; the soft-terms are then found to be
\be
\delta m^2_{Q,U,D}=-\frac{g^2_F}{16\pi^2} {|h\mu|^2}
\left(
\begin{array}{ccc}
 1.6 & 0 & 0 \\
 0 & 3.8  & 0 \\
 0 & 0 & 5.5  \\
\end{array}
\right)\, .
\ee
The stop contribution is enhanced sufficiently but the flavour mediated contribution does not approximately preserve $SU(2)_F$ symmetry. Nevertheless it has the right kind of structure, so we expect some alleviation of the naturalness problem at the expense of larger FCNCs. 

\section{Supersymmetry breaking and the low energy spectrum} \label{susybreaking}
As discussed before,  we should combine gauge mediated supersymmetry breaking in the SM gauge groups $SU(3)_c\times SU(2)_L \times U(1)_Y$ and gauge (vector multiplet) messengers in the $SU(3)_F$ group pictured in Fig.~\ref{figureofsetup}. The soft masses from $SU(3)_c\times SU(2)_L \times U(1)_Y$, are given by 
\be
M_{\lambda_i}=k_i\frac{\alpha_{i}}{4\pi} \Lambda_{\lambda_i}
\ee
\be
m_{\tilde{f}_{3\times 3}}^2=2 \sum_{i=i}^3 k_i C_i \frac{\alpha^2_{i}}{(4\pi)^2} \Lambda^2_{S_i}\left(\begin{array}{ccc}
1 & 0 & 0 \\
0 & 1 & 0 \\
0 & 0 & 1 \end{array}\right) 
\ee
where $k_i=(5/3, 1,1)$ to GUT normalise the fine-structure constants.  In the limit $\Lambda_{\lambda_i}\simeq \Lambda_S\simeq \Lambda_F$ where $\Lambda_F = F/M_{\rm mess}$, the setup will represent minimal GMSB.    We want to combine this with $SU(3)_F$  gauge messenger mediation
\be
\delta m^2_{Q,U,D}=-\left(\frac{\alpha_F}{4\pi}\right)\Lambda^2_F \mathcal{N}_{3\times 3}
\ee
where $\mathcal{N}_{3\times3}$ is a three by three matrix whose entries are a prediction of each particular model but which typically has a leading diagonal form. To generate the observed Higgs mass we assume the NMSSM. To analyse the spectrum we start from the NMSSM model file of  {\tt SARAH 4.0.3} \cite{Staub:2008uz,Staub:2010jh,Staub:2012pb,Staub:2013tta} in the super-CKM basis and implement the boundary conditions at the messenger scale. We set the model up to run the RGEs from the messenger scale $M_{\rm mess}$ down to $M_{EWSB}$.  After the {\tt SPheno}  model is built in  {\tt SARAH}  we pass it over to {\tt SPheno 3.2.4} \cite{Porod:2003um,Porod:2011nf} to carry out the running.\newline
\\
Let us now discuss the results of the low energy spectrum of the models given in this paper, namely the case of magnetic meson vev in which $\tilde{N}=2$, the two cases in which magnetic quark vevs are used to generate the Yukawa structure, and the perturbative $\tilde{N},\tilde{N}'=1$ flavour model.
\\

\begin{figure}[h!]
\begin{center}
\includegraphics[width=0.49\textwidth]{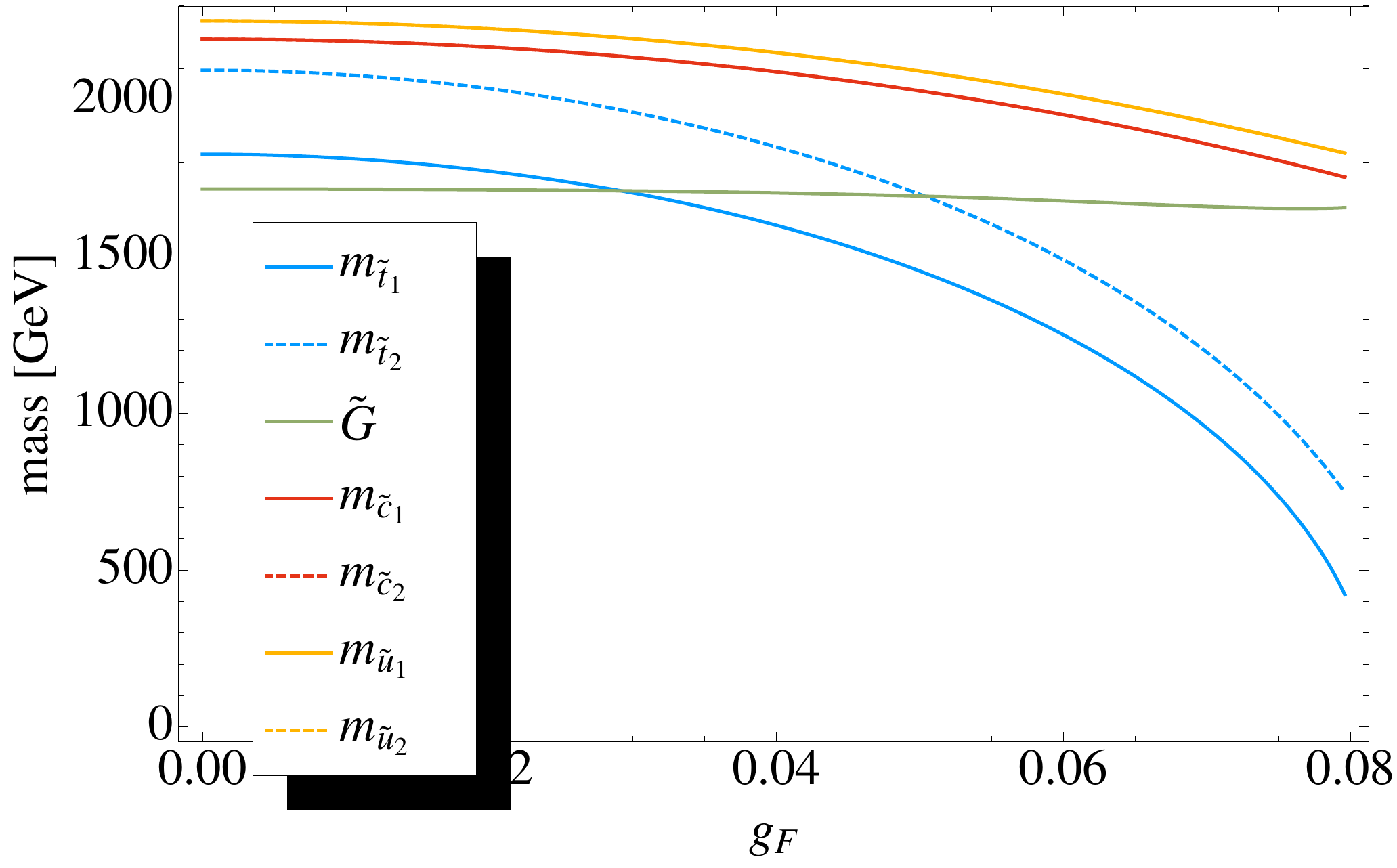}
\includegraphics[width=0.49\textwidth]{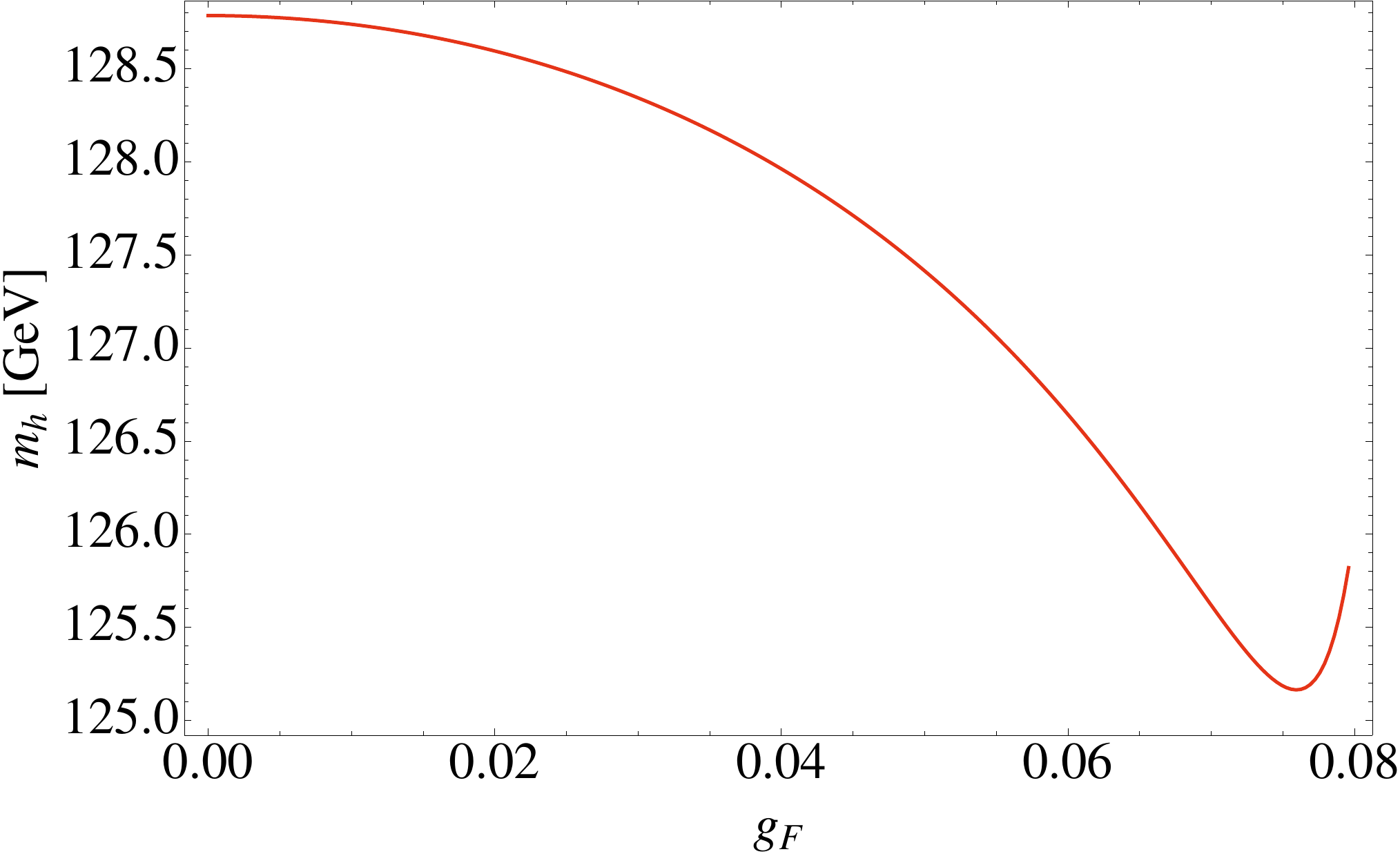}
\caption{A plot [Left] of the squark and gluino masses for model 1 with the NMSSM. [Right] a plot of Higgs mass versus $g_F$ for the same range. $\lambda=0.8$, $\kappa=0.8$, $v_s=1000$, $m_{H_d}^2=m_{H_u}^2=10^5$, $\Lambda_{\lambda_i}=\Lambda_S=\Lambda_F=2.3\times10^5$, $M_{\rm mess}=10^7$,  $\tan \beta =1.5$.} 
\label{fig:ModelA}
\end{center}
\end{figure}

\begin{figure}[h!]
\begin{center}
\includegraphics[width=0.49\textwidth]{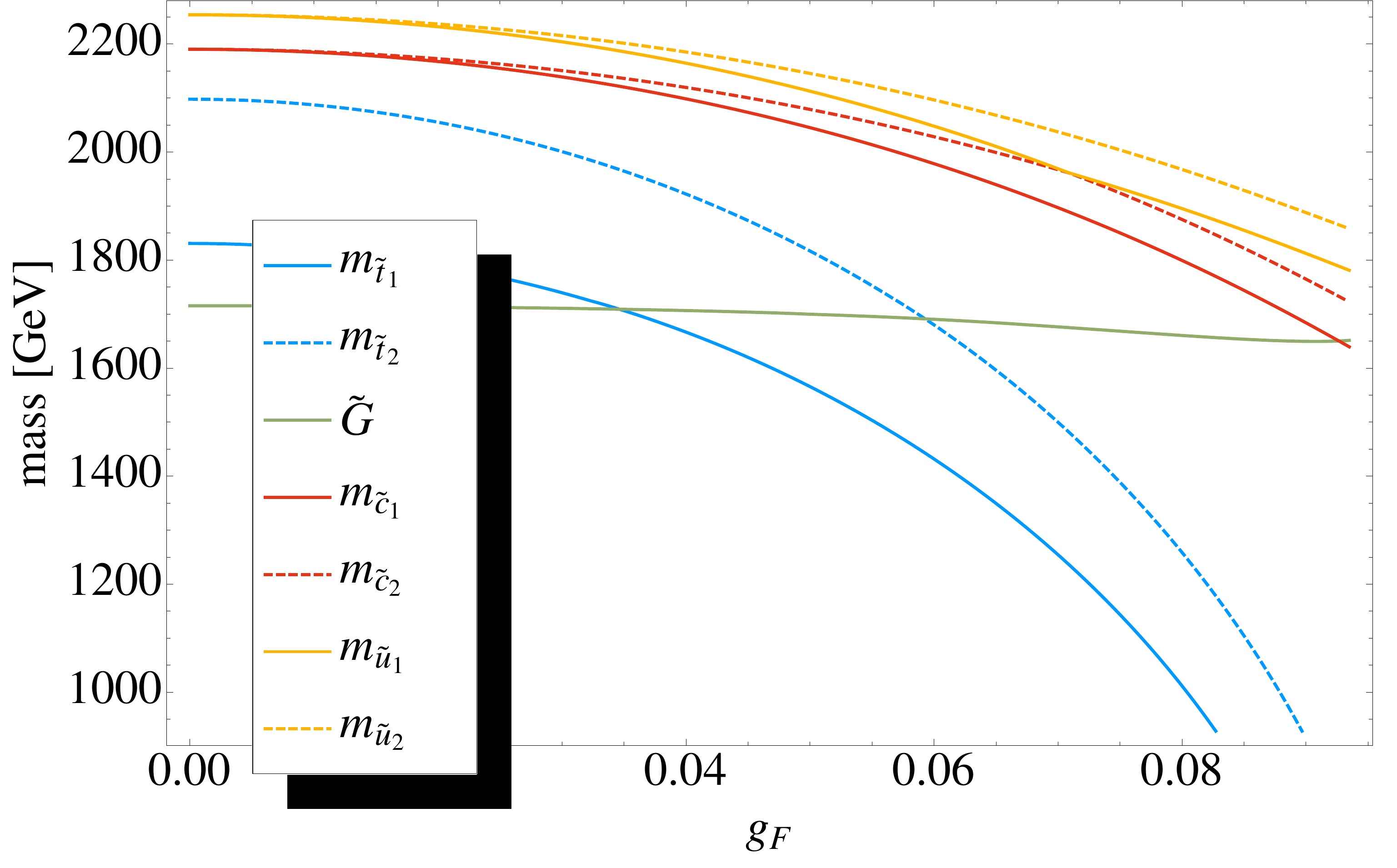}
\includegraphics[width=0.49\textwidth]{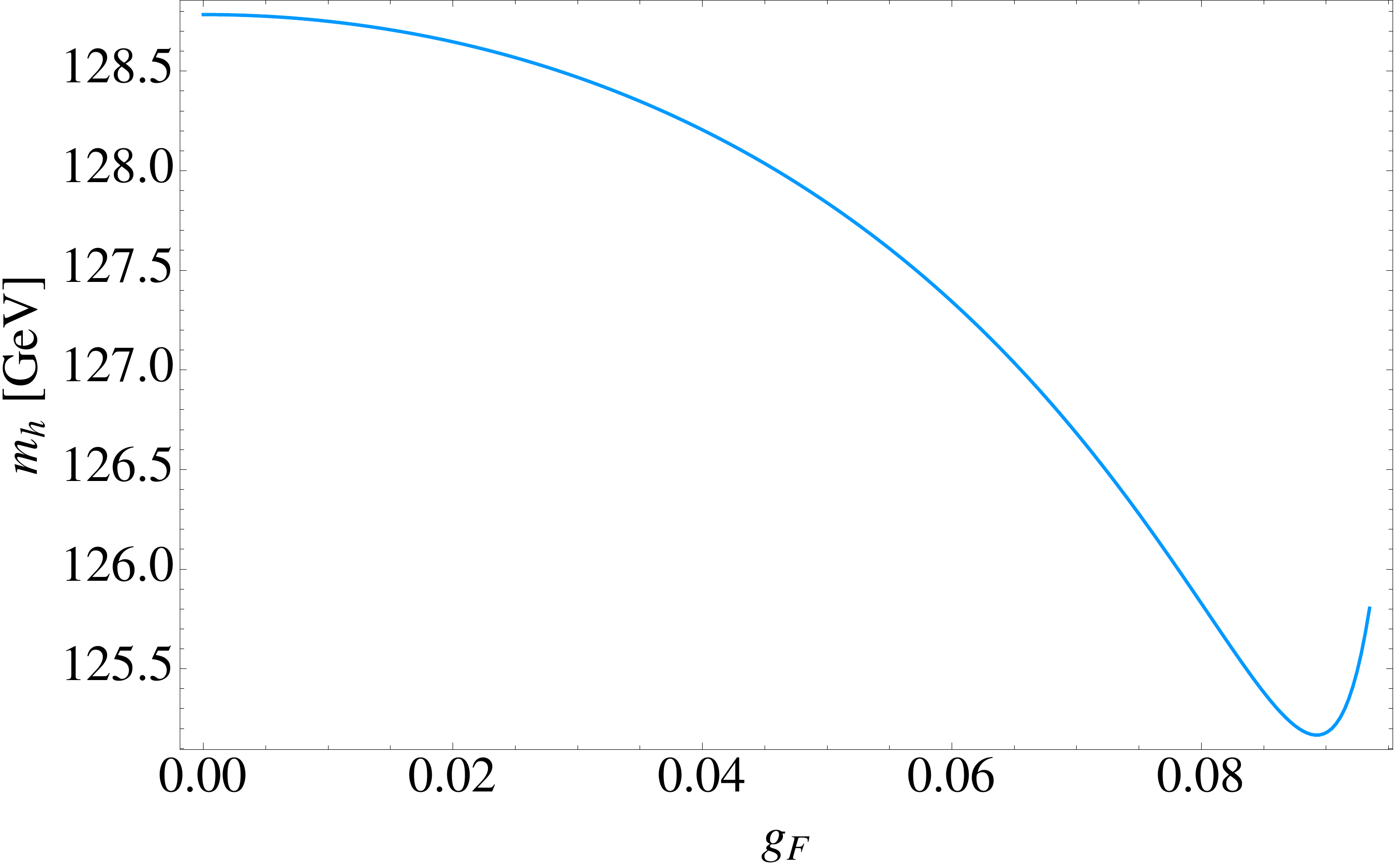}
\caption{A plot [Left] of the squark and gluino masses for model 2  with the NMSSM. [Right] a plot of Higgs mass versus $g_F$ for the same range. $\lambda=0.8$, $\kappa=0.8$, $v_s=1000$, $m_{H_d}^2=m_{H_u}^2=10^5$, $\Lambda_{\lambda_i}=\Lambda_S=\Lambda_F=2.3\times10^5$, $M_{\rm mess}=10^7$,  $\tan \beta =1.5$.} 
\label{fig:ModelB}
\end{center}
\end{figure}

\begin{figure}[h!]
\begin{center}
\includegraphics[width=0.49\textwidth]{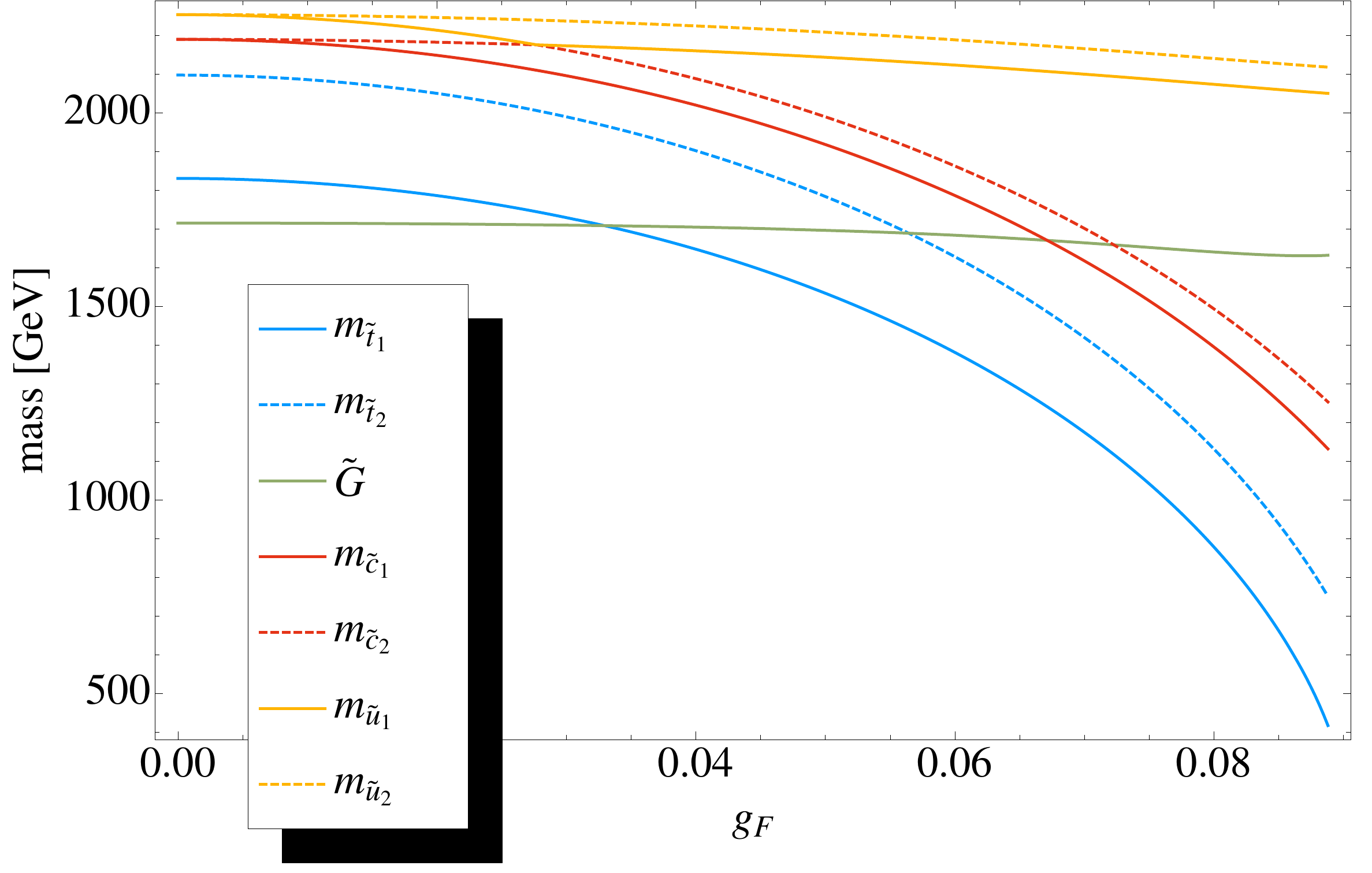}
\includegraphics[width=0.49\textwidth]{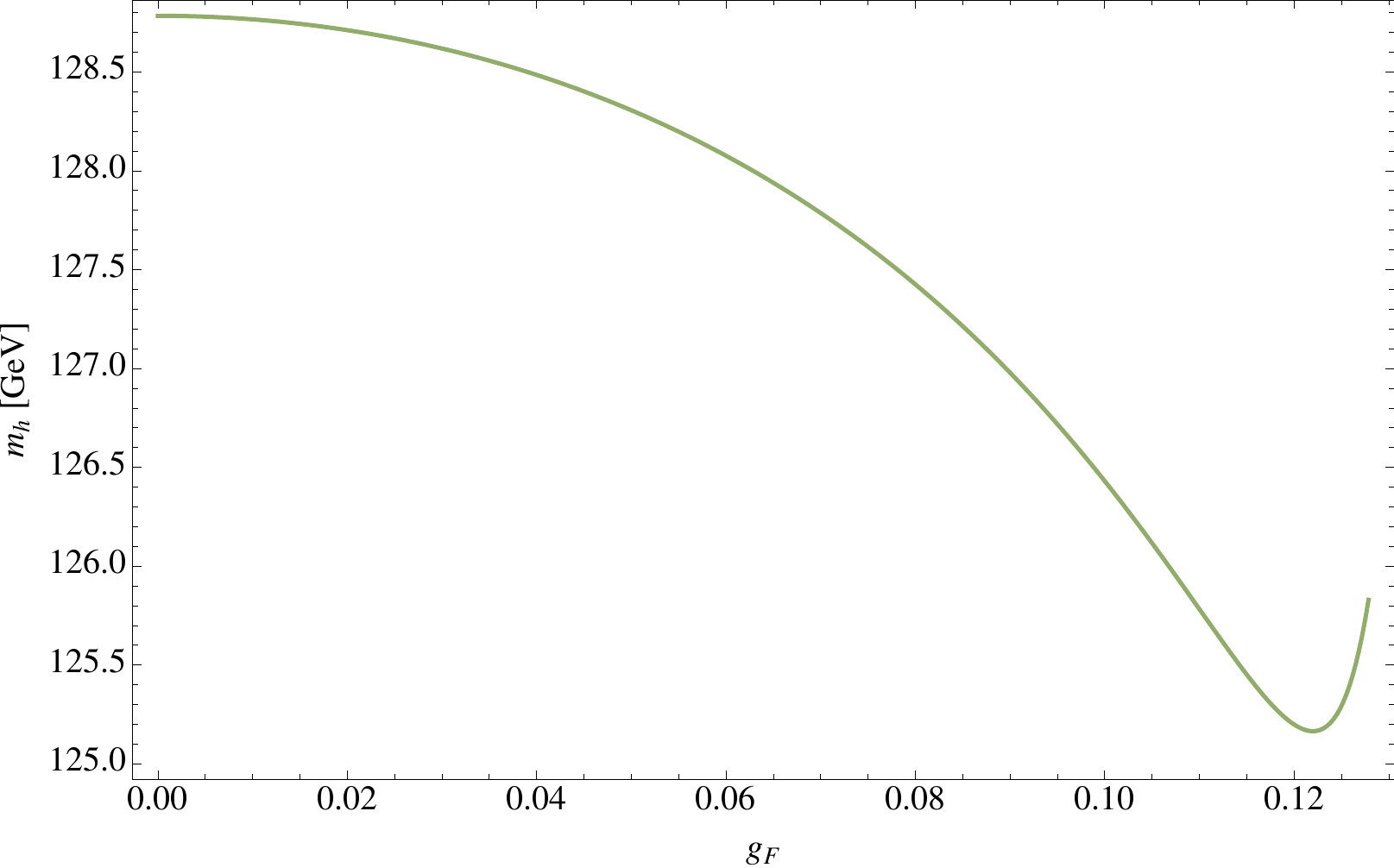}
\caption{A plot [Left] of the squark and gluino masses for model 3  with the NMSSM. [Right] a plot of Higgs mass versus $g_F$ for the same range. $\lambda=0.8$, $\kappa=0.8$, $v_s=1000$, $m_{H_d}^2=m_{H_u}^2=10^5$, $\Lambda_{\lambda_i}=\Lambda_S=\Lambda_F=2.3\times10^5$, $M_{\rm mess}=10^7$,  $\tan \beta =1.5$.} 
\label{fig:ModelC}
\end{center}
\end{figure}

\begin{figure}[h!]
\begin{center}
\includegraphics[width=0.48\textwidth]{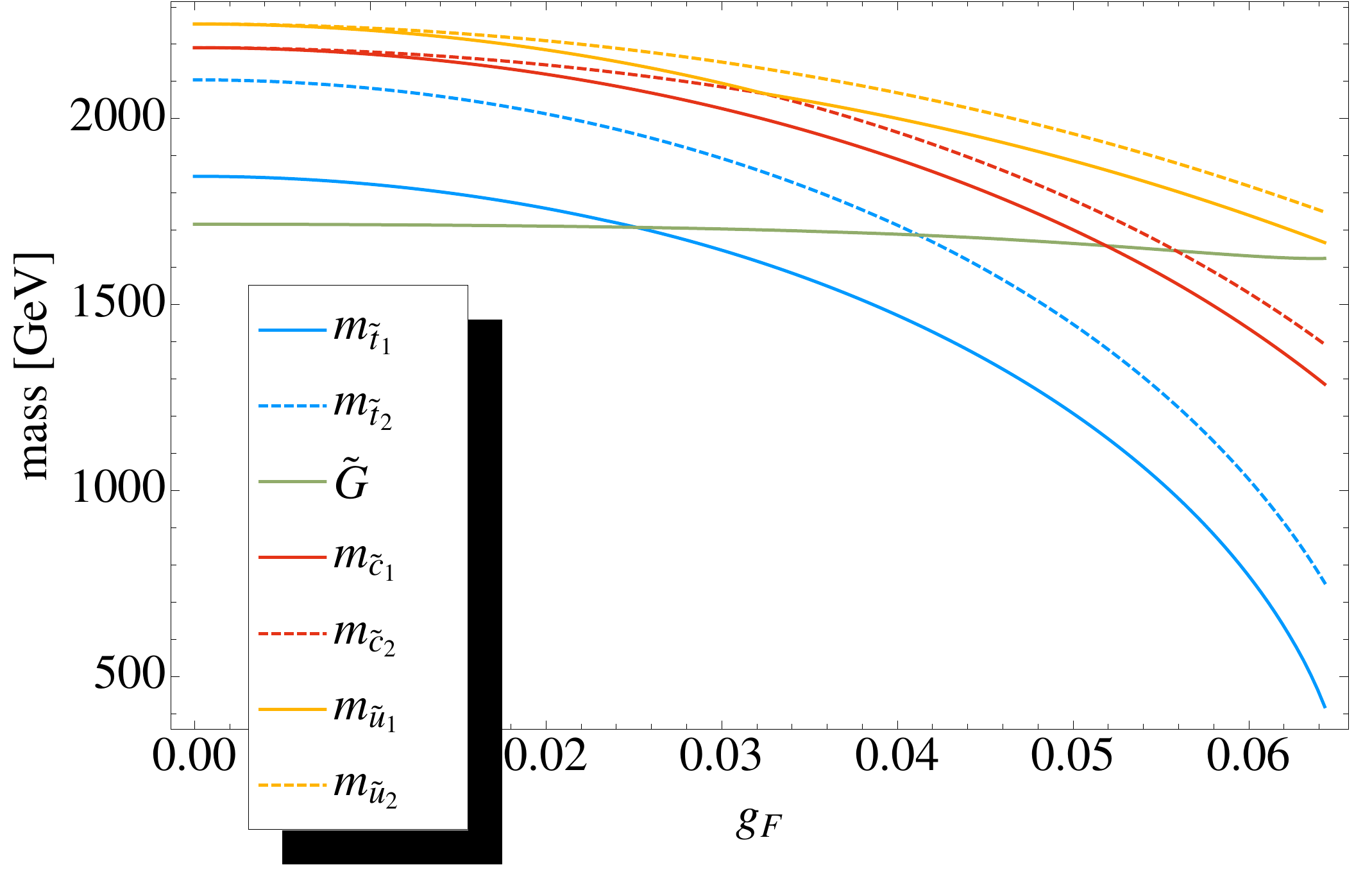}
\includegraphics[width=0.49\textwidth]{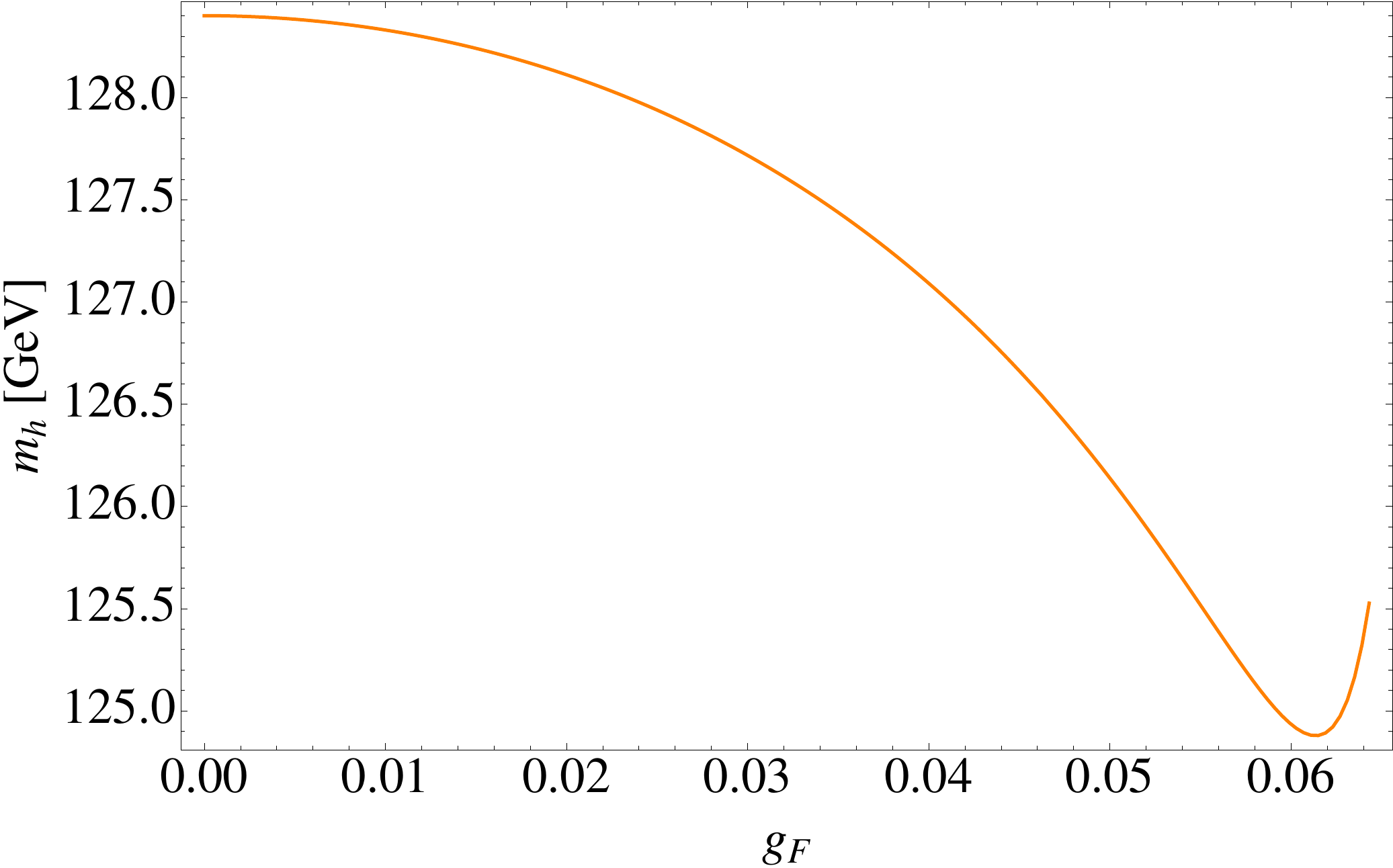}
\caption{A plot [Left] of the squark and gluino masses for the perturbative model of flavour in section \ref{perturbativemodelofflavour} with the NMSSM. [Right] a plot of Higgs mass versus $g_F$ for the same range. $\lambda=0.8$, $\kappa=0.8$, $v_s=1000$, $m_{H_d}^2=m_{H_u}^2=10^4$, $\Lambda_{\lambda_i}=\Lambda_S=\Lambda_F=2.3\times10^5$, $M=10^7$,  $\tan \beta =1.7$.} 
\label{fig:Modelsteve}
\end{center}
\end{figure}

\noindent{\bf[Model 1]}: The mesonic case. The soft masses for the flavour gauge mediation are given by \refe{eq:softmodel1} at leading order.  The results are in figure \ref{fig:ModelA}.  This model has an approximate $SU(2)$ and will have best compatibility with FCNC constraints.
\newline
\noindent{\bf[Model 2]}: The magnetic quark case with $\tilde{N}=1$.  The $SU(3)$ symmetry is completely broken resulting in non degenerate first and second generation tachyonic contributions to the soft masses.  The results are in figure \ref{fig:ModelB}. The behaviour is overall similar to model 1 but now each squark 1,2 pair are split for each generation and the effect increases with $g_{F}$ increasing.
\newline
\noindent{\bf[Model 3]}: The magnetic quark case with $\tilde{N}=2$.  As before, the $SU(3)$ symmetry is completely broken resulting in non degenerate first and second generation tachyonic contributions to the soft masses.  The results in figure \ref{fig:ModelC} show a much slower reduction in the stop mass values as $g_F$ is increased and increased splitting between first and second generation, therefore increased FCNCs are likely in this model for larger $g_F$, and this model is almost certainly excluded for this range of $\Lambda_F$.
\newline
\noindent{\bf[Model 4]}: The perturbative $\tilde{N}={\tilde{N}}'=1$ model. A typical example of the up-squark spectrum is shown in figure \ref{fig:Modelsteve}. The stop contribution is enhanced sufficiently in the spectrum to achieve the correct Higgs mass with sub TeV stops.  As the flavour mediated contribution does not approximately preserve $SU(2)_F$ symmetry, there is relatively large first-second generation splitting which means that this model would be bound to show significant deviations for FCNCs; with the heavier squark masses however this is at the margins of exclusion. (We discuss this in more detail presently).

In this class of models some squarks, usually those of the third generation, have tachyonic UV boundary conditions.   As long as $m_{\tilde{t}}$ is not larger than $M_{\tilde{g}}$ we should expect that the stops turn positive by the electroweak scale as per the figures.  It is interesting also that in 
\be
\delta m^2_{H_u}\simeq -\frac{3}{4\pi^2}m_{\tilde{t}}^2 \text{log}\frac{M_{\rm mess}}{m_{\tilde{t}}}
\ee
the contributions in the regime of negative and positive $m^2_{\tilde{t}}$ may cancel out, reducing the little hierarchy problem rather nicely \cite{Dermisek:2006ey,Dermisek:2006qj}. Larger stop mixing may be achieved at the electroweak scale also, from UV boundary conditions with no mixing ($A_t(M_{\rm mess})=0$).


\section{Constraints from flavour}\label{flavourconstraints}
The addition of flavour gauge messengers introduces some non-universal soft terms and additional massive vector states in the flavour sector.  One must ensure that these do not introduce flavour changing neutral currents.  Typically in these models the mass of the flavour gauge bosons may be computed from 
\be
(m^2_C)^{AB}=(m^2_V)^{AB}+(\delta m_0^2)^{AB},
\ee
where 
\be
(m^2_V)^{AB}=g^2(\bar{\phi},\phi)^{AB}, \ \ \  (m_{\chi})^{AB}=\frac{(\bar{\phi},F)^{AB}}{(\bar{\phi},\phi)^{AB}}, \ \ \ (\delta m_0^2)^{AB}=\frac{(\bar{F},F)^{AB}}{(\bar{\phi},\phi)^{AB}}.
\ee
Essentially these states will have mass squared roughly of the order of  $\sim g^2_F M^2$ or a little smaller. As in \cite{Craig:2012di}   integrating out the gauge bosons will give dimensions 6 operators 
\be
\mathcal{L}\supset -\frac{g_F^2}{2m_V^2}(\bar{f}^i_{L,R}\gamma^{\mu}T^a_{ij} f^j_{L,R})(\bar{f}^k_{R,L}\gamma_{\mu}T^a_{kl} f^l_{R,L}).
\ee
In order to avoid FCNCs from this source, it then sufficient to set the messenger scale to be above $\sim 10^5$ TeV  to protect against for example $K^0-\bar{K}^0$ mixing, which is the most constraining (CP preserving) $\Delta F= 2$ FCNC process. In our examples (see figures \ref{fig:ModelA},\ref{fig:ModelB} and \ref{fig:ModelC}) we have chosen $M_{\text{mess}}= m_V \sim 10^7$ GeV or higher.  

As we are introducing flavour splittings amongst the squarks, the other obvious source of flavour violation comes from radiative contributions. Our principle concern is the up (rather than down) squark sector since it is chiefly the stop that has to be made light for the model to be natural. Some of our models break the approximate $SU(2)_F$ degeneracy of the first and second generation and so it is useful to see qualitatively how non-degenerate these may be within the current flavour constraints. These bounds are usually expressed in terms of the parameter \cite{Galon:2013jba}
\be
\delta_{ij}=\frac{m_{q_2}^2-m_{q_1}^2}{\frac{1}{2}(m_{q_2}^2+m_{q_1}^2)} K_{ij} K^{*}_{ii},
\ee
where $K$ is the rotation matrix from the quark-mass basis to the squark mass-basis. As we are aligning supersymmetry breaking and flavour it is clearly possible within this context to avoid some of the bounds that may be most severe. For example if the flavour/susy breaking was aligned such that the down/sdown sector was entirely diagonal, then the most stringent FCNC constraints  ($\delta_{d,12}<0.07$ \cite{Galon:2013jba})  are avoided. Nevertheless in that case one has the constraint from the ups 
($\delta_{d,12}<0.1$ \cite{Galon:2013jba}) which is only slightly less severe. Therefore it seems reasonable to take the latter as a representative bound. It is also reasonable to suppose that the Cabibbo angle is representative of the order parameter that breaks $SU(3)_F$, so one might expect mild suppression of $K_{ij} K^{*}_{ii}\simeq 0.22$.

\begin{figure}[h!]
\begin{center}
\includegraphics[width=0.6\textwidth]{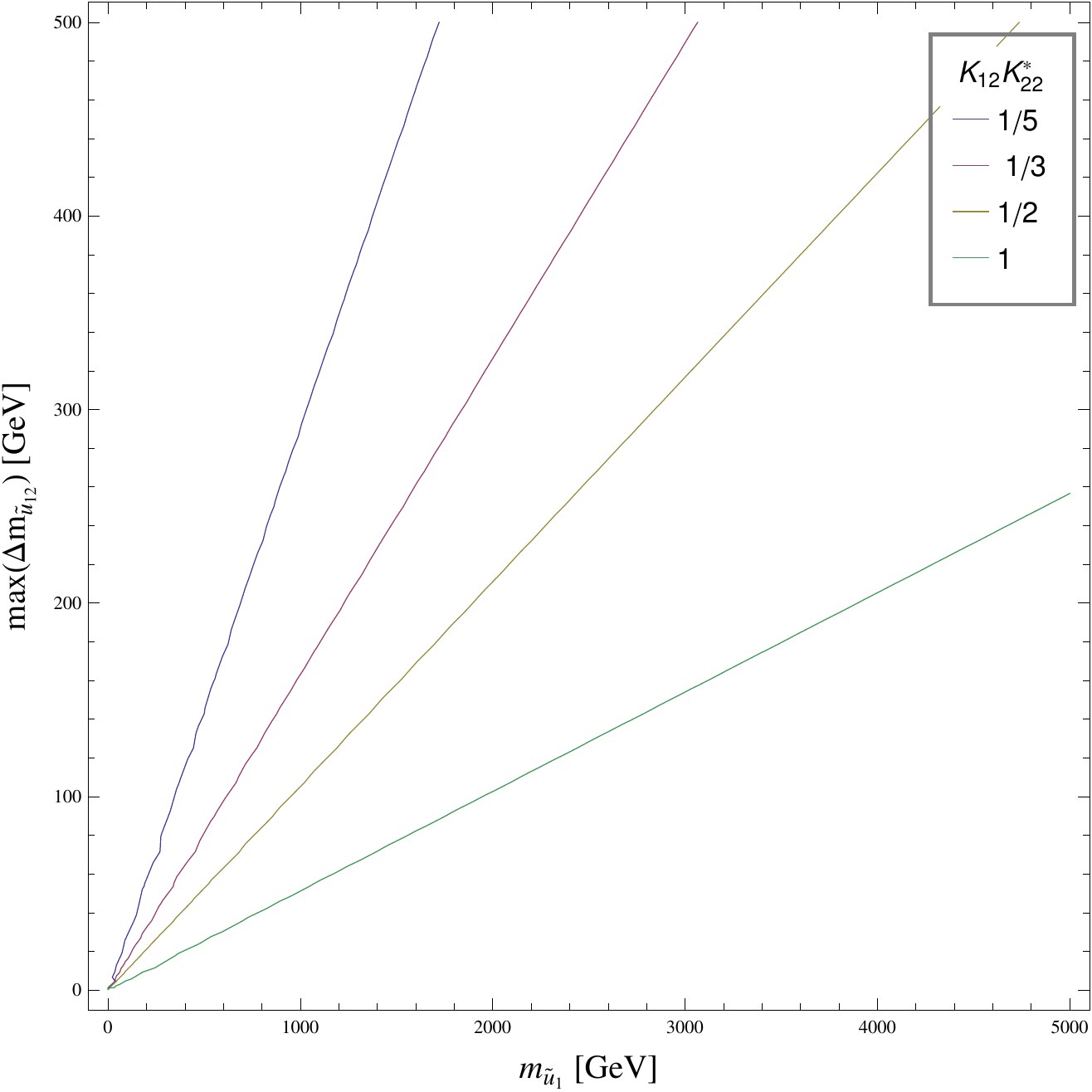}

\caption{Maximum allowed splitting between first and second generation up squarks, for a given first generation up squark, for $\delta_{12}=0.1$.} 
\label{fig:stopsplitting}
\end{center}
\end{figure}

To guide the eye, 
figure \ref{fig:stopsplitting} shows the maximum allowed splitting between first and second generation up squarks in this case for a given first generation squark mass, for various values of $K^*_{11} K_{12}$. With $\sim1.5$ TeV first and second generation one can allow as much as $400$ GeV mass-splitting if the maxing matrix $K$ has Cabibbo sized off-diagonal entries. Obviously heavier first and second generation, in the $5-10$ TeV range, allow $500$ GeV- $1$ TeV splittings.  Comparing with our models, we find  that the models 1 \& 2 in figures \ref{fig:ModelA} and \ref{fig:ModelB}  respectively are clearly allowed, whilst model 3 in figure \ref{fig:ModelC} is likely to be excluded by flavour constraints.  The perturbative flavour model in figure \ref{fig:Modelsteve} is a borderline case for such low values of $\Lambda_F$ but by increasing $\Lambda_F$  to achieve $5-10$ TeV first and second generation, the constraints from flavour may be alleviated substantially. As remarked above constraints on the down sector are slightly more stringent but the discussion is qualitatively the same.

\section{Discussion}\label{Discussion}
In this paper we have shown how various prototypes of flavour structure and supersymmetry breaking may be given a common dynamical origin in a magnetic SQCD. 
The models we presented are based on weakly gauging an $SU(3)_F$ subgroup of the SQCD global symmetry to play the role of the SM flavour symmetry, as in table \ref{tableMSSM}.  The $SU(3)_F$ is broken at the same time as supersymmetry by the ISS mechanism, resulting in  gauge (vector multiplet) messengers that generate tachyonic soft terms.  The breaking of the $SU(3)_F$ is naturally aligned with the supersymmetry breaking and these soft terms compete with the usual GMSB soft terms. This tachyonic contribution affects mainly the third generation scalars, and a realistic and natural spectrum is generated. 

The underlying description of flavour can be weakly or strongly coupled. Either way, thanks to the ISS mechanism one has a calculable and predictive framework.  Implicitly, the scale of the usual SM gauge mediation contributions and the flavour mediation contributions are assumed to be of the same order, and set by the UV physics, so as to achieve light stops.  

For reasonable phenomenology the models we consider tend to contain negative soft mass-squareds at the messenger scale for the third generation. It is worth mentioning that this need not lead to tachyons because the first two generations can have positive mass-squareds there. A tachyonic direction could potentially develop along $F$ and $D$ flat directions such as the $UDD$ direction but as described in the Introduction, these can be lifted by the positive contributions of higgs and sfermion mass-squareds along these directions. Moreover  even if the tachyonic mass-squareds do induce charge and colour breaking minima, they appear at scales close to the mediation scale which is parametrically larger than $\sqrt{F}$. As such they would have a similar status to the SUSY restoring minima of ISS. (Moreover the EWSB vacua is closer to the origin in field space with higgsed states scaling as $m^2\sim g^2\braket{\phi}^2$, so at finite temperature we would expect the system to be driven into the vacuum at the origin with more light degrees of freedom, much as in \cite{Abel:2006cr,Abel:2006my}.) 

These models are consistent with both LHC data and naturalness. 
Their ``smoking gun'' is that in contrast to minimal GMSB it does allow for FCNCs at (and violating) present limits, but avoiding them 
does not apparently require significant fine tuning at least within the limited scope of the flavour models we constructed for this analysis. It is obviously of interest to extend the analysis to complete models of flavour, to perform a more comprehensive study of these constraints. The parameter space of these models may also be excluded either by direct stop searches, which reduce the allowed values of $g^2_F \Lambda_F^2$. 
Exclusions for the stau NLSP \cite{CMS:2013jfa,ATLAS:2013ama} would also reduce the allowed values of $g^2_F \Lambda_F^2$ and thereby reduce its effect on the stop sector.   

These models are among the first that elevate the masses of the first and second generation squarks dynamically and in a purely gauge mediated framework.  This may offer a natural way to achieve stop-scharm  admixtures \cite{Blanke:2013uia} which would further alleviate collider bounds. In extending this work, it may be interesting to consider realisations based on SO(N) or $Sp(N)$ groups. 


\appendix


\bibliographystyle{JHEP}
\bibliography{flavour}

\end{document}